\documentclass[a4paper,fleqn,usenatbib, hyperref]{mnras}

\usepackage[T1]{fontenc}
\usepackage{ae,aecompl}
\usepackage{xcolor}
\usepackage[utf8]{inputenc}

\hypersetup{draft}

\usepackage{graphicx}	
\usepackage{amsmath}	
\usepackage{amssymb}	
\usepackage{pdflscape}	
\usepackage[english]{babel}

\newcommand{\mdotgas}{\dot{M}_{\rm g}}
\newcommand{\mjup}{\mathrm{M}_{\rm Jup}}
\newcommand{\mearth}{\mathrm{M}_{\rm \oplus}}

\newcommand{\rpeb}{r_{\rm peb}}
\newcommand{\rdotpeb}{\dot{r}_{\rm peb}}
\newcommand{\mdotpeb}{\dot{M}_{\rm peb}}
\newcommand{\mdotp}{\dot{M}_{\rm pl}}
\newcommand{\mdotptwod}{\dot{M}_{\rm pl,2D}}
\newcommand{\mdotpthreed}{\dot{M}_{\rm pl, 3D}}
\newcommand{\tstop}{\tau_{\rm s}}
\newcommand{\mfp}{\lambda_{\rm mfp}}
\newcommand{\rsmin}{r_{\rm str,min}}
\newcommand{\rsmax}{r_{\rm str, max}}

\title[Pebble accretion in self-gravitating discs]{Pebble accretion in self-gravitating protostellar discs}

\author[D. H. Forgan]
{D.~H.~Forgan$^{1}$\thanks{Contact e-mail: \href{mailto:dhf3@st-andrews.ac.uk}{dhf3@st-andrews.ac.uk}},
\vspace{0.2cm} \\
$^{1}$Centre for Exoplanet Science, SUPA, School of Physics \& Astronomy, University of St Andrews, St Andrews KY16 9SS, UK \\
}

\date{Accepted XXX. Received XXX; in original form XXX}

\pubyear{2018}

\begin{document}
\label{firstpage}
\pagerange{\pageref{firstpage}--\pageref{lastpage}}
\maketitle

\begin{abstract}
Pebble accretion has become a popular component to core accretion models of planet formation, and is especially relevant to the formation of compact, resonant terrestrial planetary systems.   Pebbles initially form in the inner protoplanetary disc, sweeping outwards in a radially expanding front, potentially forming planetesimals and planetary cores via migration and the streaming instability.  This pebble front appears at early times, in what is typically assumed to be a low mass disc.  We argue this picture is in conflict with the reality of young circumstellar discs, which are massive and self-gravitating.  We apply standard pebble accretion and streaming instability formulae to self-gravitating protostellar disc models. Fragments will open a gap in the pebble disc, but they will likely fail to open a gap in the gas, and continue rapid inward migration.  If this does not strongly perturb the pebble disc, our results show that disc fragments will accrete pebbles efficiently.  We find that in general the pebble-to-gas-density ratio fails to exceed 0.01, suggesting that the streaming instability will struggle to operate.  It may be possible to activate the instability if 10 cm grains are available, and spiral structures can effectively concentrate them in regions of low gravito-turbulence.  If this occurs, lunar mass cores might be assembled on timescales of a few thousand years, \emph{but this is likely to be rare, and is far from proven}.  In any case, this work highlights the need for study of how self-gravitating protostellar discs define the distribution and properties of solid bodies, for future planet formation by core accretion.

\end{abstract}




\section{Introduction}
\label{sec:introduction}

\noindent Terrestrial planet formation requires the growth of cores from small dust grains.  This is true regardless of which planet formation mechanism we consider.  If planet formation proceeds via the core accretion paradigm, then 100 - 1000 km planetesimals are assembled in the protoplanetary disc, which can then grow into cores, with accretion of a gaseous atmosphere proceeding once the core has reached sufficient mass \citep[cf][]{Pollack1996,Mordasini2008}.  If planet formation proceeds via the gravitational instability (GI) and disc fragmentation, then each disc fragment, with mass approximately a few Jupiter masses \citep{Boley2010b,Forgan2011a,Kratter2011} contains a subsample of the disc's grain population.  This subsample of grains can also grow and settle to the centre of the fragment to form a core \citep{Helled2008,Helled2008a}.  The eventual fate of the fragment depends on the ordering of multiple timescales - the inward migration timescale, which determines when the fragment is affected by tidal disruption from the central star; and the core formation timescale, which is itself a function of the grain growth and sedimentation timescales.

Tidal downsizing theory addresses the multiple products of this timescale ordering \citep[see][for a review]{Nayakshin2017}.    Most orderings result in massive gas giants or brown dwarfs.  If migration is sufficiently slow, and core formation sufficiently rapid, the fragment's gas can be removed by tidal forces or photoevaporation to leave a terrestrial planet.  Our work indicates that terrestrial planet formation by tidal downsizing is rare \citep{TD_synthesis,TD_nbody}.  That being said, there is a growing body of evidence that planet formation occurs early, while the disc is still young and relatively massive.  Some examples include the gaps observed in the disc of HL Tau \citep{Brogan2015,Dipierro2015a,Carrasco-Gonzalez2016,Boley2017}, the spiral structure in the Elias 2-27 system, consistent with the presence of a companion \citep{Perez2016,Meru2017,Forgan2018b}, and the gaps observed in the $\sim$ 0.4 Myr old Elias 24 system, consistent with a $\sim 0.7 \mjup$ planet present  \citep{Dipierro2018}.

Both planet formation mechanisms have experienced difficulties in explaining some of the more exotic exoplanet systems observed.  A long-standing problem for core accretion theory has been the so-called radial drift barrier or metre barrier \citep{Whipple1973,Weidenschilling1977}.  Interstellar dust grains (of order a few $\micron$ in size or smaller) are very tightly coupled to the local disc gas.  Equivalently, the stopping time is much shorter than the local dynamical timescale, which we can represent using the dimensionless stopping time (or Stokes number) $\tau_s$ as

\begin{equation}
\tau_s = t_{\rm stop} \Omega <<1
\end{equation}

\noindent As grains grow in the disc, $\tau_s$ approaches unity and the grain starts to feel significant aerodynamic drag as it attempts to move at Keplerian angular velocity in a gas disc that is sub-Keplerian due to pressure effects.  Grains experiencing this headwind begin to drift inwardly, with the drift timescale becoming increasingly rapid as the grains reach metre size (the peak of the drift timescale being somewhat dependent on the local disc properties).  This rapid loss of grains onto the star is compounded by the fragmentation and bouncing barriers, which prevent grain collisions from resulting in successful accretion \citep{Guttler2010,Windmark2012}.  Against these combined challenges, the formation of planetesimals from these grains undergoing heavy drag (which we refer to as \emph{pebbles}) appeared essentially impossible.  Some models have showed means of overcoming these barriers by considering the particle velocity distribution in more detail \citep{Windmark2012b,Garaud2013,Booth2018}, but these are yet to be implemented in global disc models with appropriate radial drift.

In recent years, two physical processes have been described that appear to assist the core accretion process.  The first is a recognition that the aerodynamic drag felt by the grains creates a back-reaction on the gas, increasing the local gas velocity.  If grains can sufficiently concentrate into clusters, the local headwind they experience can be reduced, reducing the inward drift velocity of the cluster. As further grains are accreted into the cluster, the headwind is reduced further and the cluster can grow exponentially - the \emph{streaming instability} \citep{Youdin2005,Johansen2007,Johansen2007b, Simon2016}.  Once the instability is activated, this can result in the rapid assembly of planetesimals, which go on to form protoplanetary cores.  The streaming instability requires a sufficient local concentration of solids, i.e. a relatively large local solid-to-gas density ratio (that can be maintained against disruption mechanisms such as turbulence) as well as aerodynamically well-coupled grains - $\tau_s \sim 0.001-0.1$ \citep{Krijt2016,Yang2017}\footnote{We should note that the criteria for the streaming instability in turbulent discs are not well understood.  For example, \citet{Youdin2005} neglect turbulence, and vertical gravitational settling in their initial calculations.  It is even less well understood how the streaming instability operates when the disc is self-gravitating.}.

These protoplanets are then embedded in the radially inward flow of pebbles.  The combination of gravitational focusing by the cores and aerodynamic drag from the disc can then act in concert to significantly increase the accretion cross-section of protoplanets, resulting in the efficient growth of rocky cores by \emph{pebble accretion}.  This can allow gas giants to form on shorter timescales than previously considered \citep{Lambrechts2012}, as their cores form more rapidly, allowing them to accrete their gas from the disc before it is dispersed.  Equally, such rapid accretion can result in atmospheric ablation of the pebbles, which can limit the final core mass as the pebbles evaporate \citep{Brouwers2018}.

It is quite common for core accretion models of planet formation to assume that pebble formation begins at an early phase in the disc's life.  Pebbles form on timescales comparable to the local dynamical timescale.  Therefore pebble formation occurs initially in the inner disc, which rapidly becomes an outwardly expanding front of inwardly migrating material.  Depending on the configuration of the disc system and the location of accreting embryos, this pebble front can become streaming unstable, and begin to form planetesimals and terrestrial planets almost immediately.  Such models can successfully build compact terrestrial systems via core accretion at relatively high efficiency - an example being the TRAPPIST-1 system \citep{Ormel2017}.  

What is not often considered in these models is what the disc actually looks like at these early epochs.  It is common for core accretion models to assume a relatively low mass disc, with angular momentum transport modelled as a turbulent viscosity with constant Shakura-Sunyaev viscosity parameter $\alpha$: 

\begin{equation}
\nu = \alpha c_s H,
\end{equation}

\noindent consistent with angular momentum transport being dominated either by the magnetorotational instability (MRI) or some other form of instability.  However, pebble formation is expected to commence only a handful of inner orbital timescales after the system's birth, with the pebble front having reached the outer disc after a few hundred thousand years \citep{Krijt2016,Ormel2017}.  This is consistent with observations of Class 0/I protostars, that suggest that the initial growth of grains to mm sizes must be rapid \citep[e.g.][]{Harsono2018}.  During this period, the disc is still massive compared to its host star.  We therefore expect the disc to be prone to the gravitational instability, which should dominate its evolution.  Indeed, we expect some discs to be sufficiently unstable to fragment into bound objects, which will eventually encounter this pebble front as it sweeps outward.  Even without disc fragmentation, the physics of gravitational instability must have profound effects on the mass and spatial distribution of dust grains \citep{Gibbons2012,Booth2016,Nixon2018}.

In this paper, I ask: what does the current crop of pebble accretion models for core accretion predict/imply for young self-gravitating protostellar discs? More specifically, I ask two questions:

\begin{enumerate}
\item Can self-gravitating disc fragments accrete pebbles efficiently?
\item Can terrestrial mass planets be formed in self-gravitating discs without the aid of disc fragments, via pebble fronts and the streaming instability?
\end{enumerate}

In short, \emph{are the core accretion and gravitational instability theories of planet formation indelibly linked by pebble accretion?}

I approach these questions by applying the pebble front model of \citet{Ormel2017} to steady state self-gravitating disc models, to compute the effect of pebble accretion on disc fragments, and to determine whether planetesimals (and terrestrial planet cores) can be formed rapidly by the streaming instability during the self-gravitating phase.  Section \ref{sec:method} outlines the model construction; section \ref{sec:results} describes the results and section \ref{sec:discussion} discusses their implications for both planet formation mechanisms; section \ref{sec:conclusions} summarises the work.


\section{Method}
\label{sec:method}

\subsection{Quasi-steady self-gravitating disc models}

\noindent We construct marginally stable self-gravitating disc models in the same manner as \citet{Forgan2011a,Forgan2013a,Forgan2016e} \citep[see also][]{Rafikov_05,Clarke_09}. These steady-state models are defined by a fixed value of the Toomre Parameter \citep{Toomre_1964}:

\begin{equation}
Q = \frac{c_s \kappa_{\rm ep}}{\pi G \Sigma_g} \sim 1,
\end{equation}

\noindent where $c_s$ is the gas sound speed, $\Sigma_g$ is the gas disc surface mass density and $\kappa_{\rm ep}$ is the epicyclic frequency (which in Keplerian discs is equal to $\Omega$).  We compute $\Omega$ assuming the total enclosed mass of the system, as opposed to just the star mass:

\begin{equation}
\Omega = \sqrt{\frac{GM_{\rm enc}(r)}{r^3}}.
\end{equation}

\noindent We make the pseudo-viscous approximation, where we can treat the gravitational instability as a turbulent viscosity acting on the disc.  The turbulent viscosity $\nu$ is \citep{Shakura_Sunyaev_73}:

\begin{equation}
\nu = \alpha c_s H_g,
\end{equation}

\noindent where $H_g=\frac{c_s}{\Omega}$ is the gas disc scale height, and $\alpha<1$ is a dimensionless parameter that relates to the local stresses generated by the instability.  We are permitted to make this approximation when the angular momentum transport is locally determined.   This is the case when the disc aspect ratio $H/r \lessapprox 0.1$, or equivalently the disc-to-star mass ratio $q<0.5$ \citep{Lodato_and_Rice_04,Lodato2005,Forgan2011}.  The value of $\alpha$ is computed assuming the disc is in local thermodynamic balance \citep{Gammie}:

\begin{equation}
\alpha = \frac{4}{9\gamma(\gamma-1)\beta_c}
\end{equation}

\noindent where $\gamma$ is the (3D) ratio of specific heats, which we set at 5/3.  We can determine the dimensionless cooling parameter

\begin{equation}
\beta_c = t_{\rm cool} \Omega = \frac{u}{\dot{u}}\Omega
\end{equation}

\noindent where $u$ is the internal energy density. By computing the radiative cooling

\begin{equation}
\dot{u} = \frac{\sigma_{SB} T^4}{\tau + \tau^{-1}}
\end{equation}

\noindent With $\sigma_{SB}$ the Stefan-Boltzmann constant, and $\tau = \Sigma_g \kappa$ is the optical depth.  We use the denominator in this form to interpolate smoothly between optically thick and optically thin cooling.  Stellar irradiation is not considered in these models.  As we are in the steady state, we assume that the disc accretion rate is constant as a function of radius:

\begin{equation}
\mdotgas = 3\pi \nu \Sigma_g = 3\pi \frac{\alpha c^2_s \Sigma_g}{\Omega} = const.
\end{equation}

\noindent where typically $\alpha\sim 10^{-2}$. These equations are sufficient to close the system, provided that we have an equation of state that allows $u$, $\kappa$ and $\gamma$ to be computed as a function of density and temperature.  Our equation of state uses \citet{Bell_and_Lin} opacities, along with an energy equation taken from \citet{Stam_2007} \citep[see their equation (37), and see also][]{Black}.  This energy equation assumes hydrogen and helium mass fractions of 0.7 and 0.3 respectively, and a fixed ratio of ortho- to-para-hydrogen of 3:1. The internal energy density of the gas is the sum of the energy densities of H and He in their various ionisation states.  The ionisation state of both H and He are pre-computed using Saha equations, assuming that the ionisation of H completes before the ionisation of He begins, and that He$^+$ ionisation is complete before He$^{++}$ ionisation begins (see \citealt{Stam_2007} for more details).

We use these combined constraints to iterate on $\Sigma_g$ at a given radius.  Repeated iterations produce a 1D disc profile $\Sigma_g(r)$ for a fixed accretion rate $\mdotgas$, out to a given outer radius.  The maximum outer radius calculated in this work is 100 au.

\subsection{Disc fragmentation}

\noindent Once a disc profile is established, it can be tested to determine if it is prone to disc fragmentation.  We use the Jeans mass formalism for determining disc fragmentation, which relies on computing the time derivative of the Jeans mass inside a spiral arm $M_J$:

\begin{equation}
M_J = \frac{4\sqrt{2Q} \pi^3}{3G} \frac{c^2_s H_g}{1+\frac{\Delta \Sigma_g}{\Sigma_g}}.
\end{equation}

\noindent The above formula (and its time derivative) are derived in \citep{Forgan2011a}. We compute the specific gas surface density perturbation amplitude of the spiral arm, $\Delta \Sigma_g/\Sigma_g$, using the empirical relation established by \citet{Rice2011}.  Their simulations indicate the time-averaged perturbation amplitude

\begin{equation}
\left<\frac{\Delta \Sigma_g}{\Sigma_g}\right> = 4.47 \sqrt{\alpha}.
\end{equation}

\begin{figure}
\begin{center}
\includegraphics[width=0.5\textwidth]{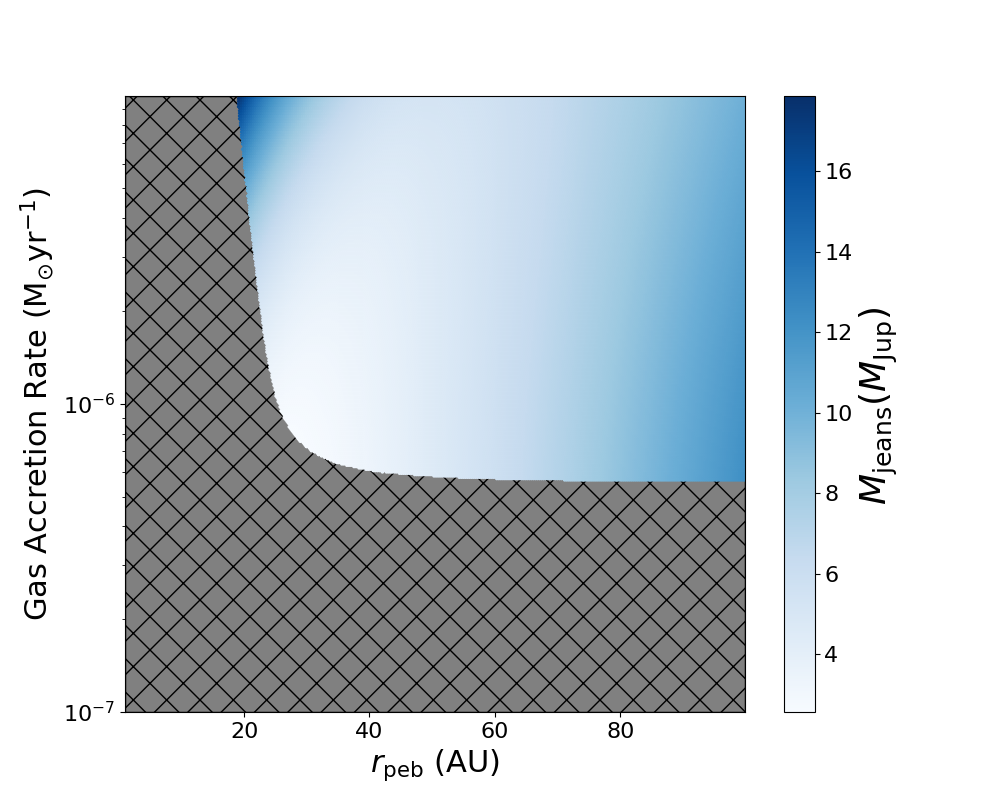}
\end{center}
\caption{Fragmentation in the self-gravitating disc model.  The upper right region of the plot indicates disc parameters that result in fragmentation, with the colour indicating the local Jeans mass at fragmentation (in Jupiter masses).  Hashed regions of the plot indicate no fragmentation.  The minimum fragment mass is approximately 3 $\mjup$. \label{fig:mjeans}}
\end{figure}

\noindent Regions of the disc are prone to fragmentation if $\dot{M}_J$ is large and negative.  Consequently, surface density perturbations generated by the spiral arms become Jeans unstable and collapse into disc fragments.  We can therefore also compute the initial mass of the fragment, which will be equal to $M_{\rm frag} = M_J$ (see Figure \ref{fig:mjeans}).

\subsection{Pebble Accretion}

\noindent With a grid of self-gravitating disc models computed, we can now use them as a backdrop on which to apply typical pebble accretion calculations.  We assume that drifting pebbles grow from small grains on a timescale given by

\begin{equation}
t_{\rm peb} = \beta_{\rm peb} \left(Z_0 \Omega\right)^{-1},
\end{equation} 

\noindent where $\beta_{\rm peb} \approx 10$ \citep{Krijt2016}, and $Z_0=0.02$.  By setting $t=t_{\rm peb}$, this defines an expression for a radially advancing front of pebbles $\rpeb(t)$.  This front generates an inward accretion flow of pebbles, with accretion rate

\begin{equation}
\mdotpeb = 2 \pi \rpeb \rdotpeb Z_0 f_{\rm peb} \Sigma_g(\rpeb).
\end{equation}

\noindent We assume throughout that the fraction of the disc's solids at pebble sizes, $f_{\rm peb}$, is 0.1\footnote{This is a relatively conservative assumption, as a large fraction of the disc's solids should reside at sizes near the fragmentation barrier.  For example, \citet{Ormel2017} set $f_{\rm peb}=0.5$}.  We fix the maximum grain size $s=1$ mm in our calculations, where this choice is motivated principally by the disc's relative youth, and our knowledge that such young discs can possess grains of at least this size \citep{Harsono2018}.  This fixes the dimensionless stopping time $\tstop$ as follows:

\begin{equation}
\tstop = \begin{cases}
		         \dfrac{\sqrt{2\pi} \rho_s s}{\Sigma_g},  & s < \dfrac{9 \mfp}{4}, \\
			     \dfrac{4\sqrt{2\pi} \rho_s s^2}{9\mfp \Sigma_g}, & s \geq \dfrac{9 \mfp}{4},
			\end{cases}
\end{equation}

\noindent where the mean free path 

\begin{equation}
\mfp = \frac{\sqrt{2\pi}\mu m_H H_g}{\sigma_{H_2} \Sigma_g}.
\end{equation}

\noindent Here, $\mu$ is the mean molecular weight of the gas, $m_H$ is the mass of the hydrogen atom and $\sigma_{H_2}$ is the collisional cross-section of $H_2$.  Throughout, we demand that the stopping time does not exceed the maximum time stopping time permitted by pebble fragmentation \citep{Birnstiel2009}:

\begin{equation}
\tau_{\rm s,f} = \frac{v^2_f}{\alpha c_s^2},
\end{equation}

\noindent where the fragmentation velocity $v_f$ is empirically determined to be 10 ms$^{-1}$ for ice-rich particles \citep{Wada2009,Gundlach2014}.  As a result, the local grain size in the disc is the minimum of $[1\,\rm{mm}, s_{\rm frag}]$ where $s_{\rm frag}$ is the grain size that yields the stopping time $\tau_{\rm s,f}$. This limits the maximum allowed grain size without pebble collisions resulting in fragmentation.

\subsubsection{Pebble accretion by disc fragments}

\noindent Wherever in the disc a fragment can be made (mass $M_{\rm frag}$), we now ask: what is the accretion rate of pebbles onto the fragment?  We employ the pebble accretion calculations of \citet{Ida2016}.  The accretion rate of pebbles onto a body is given by 

\begin{equation}
\mdotp = \rm{min} \left(\mdotptwod,\mdotpthreed\right),
\end{equation}

\noindent where the two accretion rates $(\mdotptwod,\mdotpthreed)$ are computed depending on whether the pebble flow is effectively two-dimensional (i.e. if the scale height of the pebble disc $H_{\rm peb}$ is small compared to the collision cross-section, $2b$), or fully three-dimensional.  This can be reconfigured into

\begin{equation}
\mdotp = \rm{min} \left(1, \sqrt{\frac{8}{\pi}} \frac{H_{\rm peb}}{b}\right) \sqrt{\frac{\pi}{2}} \frac{b^2}{H_{\rm peb}} \Sigma_{\rm peb} \Delta v. \label{eq:mdotp}
\end{equation}

\noindent The scale height of the pebbles is related to the gas scale height via \citep{Dubrulle1995}:

\begin{equation}
\frac{H_{\rm peb}}{H_g} =  \sqrt{\frac{\alpha}{\alpha + \tstop}} \label{eq:Hratio}
\end{equation}

\noindent The mass flow of pebbles relative to the fragment is \citep{Ida2016}:

\begin{equation}
\Sigma_{\rm peb} \Delta v = \frac{\mdotpeb}{4\pi r \tstop} \frac{\chi}{\zeta} \left(1 + \frac{3b}{2\chi \eta r}\right), \label{eq:sigma_deltav}
\end{equation}

\noindent where 

\begin{eqnarray}
\chi = \frac{\sqrt{1+4\tstop^2}}{1+\tstop^2}, \\
\zeta = \frac{1}{1+\tstop^2},
\end{eqnarray}

\noindent and $\eta$ describes the radial pressure gradient

\begin{equation}
\eta = \frac{H^2_g}{2} \left| \frac{d \ln P}{d \ln r}\right|.
\end{equation}

\noindent The cross-section parameter $b$ is given by

\begin{equation}
b = \rm{min} \left(1, \sqrt{\frac{3}{\chi \eta}\left(\frac{\tstop M_{\rm frag}}{3M_*}\right)^{1/3}}\right) R_H,
\end{equation}

\noindent where $R_H$ is the Hill radius of the fragment:

\begin{equation}
R_H = a_p \left(\frac{M_{\rm frag}}{3M_*}\right)^{1/3}.
\end{equation}

\noindent Hence, once the pebble front reaches the fragment's semimajor axis $a_p$, we can compute the pebble accretion rate by setting $\rpeb=a_p$.  We demand throughout that $\mdotp$ does not exceed the local pebble accretion flow $\mdotpeb$.


\subsubsection{Gap Opening Criteria}

\noindent Whether or not disc fragments can accrete pebbles is entirely dependent on whether the fragment opens a gap in the pebble disc.  It is already well noted in the literature that gap formation in the gaseous component of GI discs is non-trivial (see e.g. \citealt{Malik2015a}).

For the gas, we can apply two criteria to check for gap opening.  Firstly, we apply the torque balance criterion for gap opening from \citet{Crida2006}:

\begin{equation}
\Gamma_{\rm balance} = \frac{3 H_{\rm peb}}{4R_H} + 50 \alpha \left(\frac{H_{\rm peb}}{r}\right)^2 \frac{M_*}{M_{\rm frag}} \lesssim 1 \label{eq:gap_torque_balance}
\end{equation}
 
\noindent We also demand that the fragment migrates sufficiently slowly to open the gap \citep{Malik2015a}.  Formally, this requires the gap opening timescale

\begin{equation}
t_{\rm gap} = \left(\frac{H_g}{r}\right)^5 \left(\frac{M_*}{M_{\rm frag}}\right)^2 \frac{1}{\Omega} \label{eq:gap_opening_timescale}
\end{equation}

\noindent be shorter than the timescale to cross the horseshoe region (approximately $2.5 R_H$ in extent):

\begin{equation}
t_{\rm cross} = \frac{2.5 R_H}{v_{\rm mig}} = \frac{2.5 R_H t_{I}}{r}.
\end{equation}
  
But what about the pebble disc? Without exception, our disc fragments exceed the local pebble isolation mass (see equation (\ref{eq:pebbleisolationmass}) in the following section), and should therefore open a gap in the pebbles, even if they cannot open a gap in the gas disc.  

If a gap only exists in the pebble disc and not the gas, the fragment will continue inward Type I migration.  We therefore assume that the pebble accretion rate now depends on the velocity difference between the migrating planet and the pebble drift, i.e. that there is no further back-reaction on the pebble disc due to the fragment's migration.  This modifies the mass flow of pebbles relative to the fragment as follows:

\begin{equation}
\Sigma_{\rm peb} \Delta v \rightarrow \Sigma_{\rm peb} \left| v_{\rm mig} - \Delta v\right| = \Sigma_{\rm peb} \left| \frac{r}{t_{I}} - \Delta v\right|
\end{equation} 

\noindent and the pebble accretion rate is hence modified by

\begin{equation}
\mdotp \rightarrow \mdotp \left|\frac{r}{t_{I} \Delta v} - 1\right|
\end{equation}

\noindent We further assume that in the limit of large $t_{I}$, the pebble accretion rate tends to zero, although this is not achieved by any fragment in practice.

However, the pebble isolation mass is derived assuming a much lower mass disc, and it is not immediately clear that it holds for a self-gravitating disc.  At best, we can consider the effect of applying it versus not applying it.  For the parameter space explored in this work, gap opening in the pebble disc reduces the pebble accretion rate by up to a factor of approximately 5\footnote{We should also note that the equation for $t_{I}$ is likely an overestimate of the Type I migration timescale in self-gravitating discs \citep{Baruteau2011}.  Whether this allows for a boost of the pebble accretion rate, or a stronger pressure wake preventing pebbles from accreting onto the fragment is beyond the scope of this work.}.

\subsubsection{Streaming instability and rapid growth of pebbles into cores}

\noindent While there are regions of the disc not prone to fragmentation, it is worth asking if they are prone to the streaming instability.  This is likely to occur wherever the local density of pebbles exceeds the gas density in the midplane:

\begin{equation}
\left(\frac{\rho_p}{\rho_g}\right)_{\rm midplane} = \frac{\Sigma_{\rm peb}}{\Sigma_g} \frac{H_g}{H_{\rm peb}}
\end{equation}

\noindent We can use equation (\ref{eq:Hratio}) to compute the scale height ratios, and we can estimate the surface density of pebbles via

\begin{equation}
\Sigma_{\rm peb} = \frac{\mdotpeb}{2 \pi r v_r}
\end{equation}

\noindent Where the radial velocity of the pebbles is a combination of their drift velocity and the gas radial velocity \citep{Takeuchi2002}:

\begin{equation}
\left|v_r\right| =  \left|\frac{-3}{r\Sigma_g} \frac{\partial}{\partial r} \left(r^{1/2} \nu \Sigma_g\right) + 2\eta \Omega r  \frac{\tstop}{1+\tstop^2}  \right|,
\end{equation}

\noindent where the left hand term describes the viscous radial velocity, and the right the drift velocity. We can therefore identify minimum and maximum radii $(\rsmin,\rsmax)$ where the streaming instability can operate.  This streaming unstable region rapidly assembles planetesimals,.  We follow \citet{Ormel2017} in assuming that a single planetary core emerges out of this planetesimal \& pebble population. This core migrates from $\rsmax$ to a given radius $r$, continuing to accrete and grow.  We can therefore estimate a crossing mass as the core traverses the streaming unstable zone, assuming Type I migration.  

The core (mass $M_{\rm c}$) grows on the following timescale:

\begin{equation}
t_{\rm grow} = \frac{M_{\rm c}}{\mdotpeb \epsilon_{PA}} \label{eq:tgrow}
\end{equation}

\noindent Where we introduce a pebble accretion efficiency $\epsilon_{PA}$, which we leave fixed at 0.1 \citep{Liu2018,Ormel2018}.  The efficiency is defined as

\begin{equation}
\epsilon_{PA} = \frac{\mdotp}{\mdotpeb}
\end{equation}

\noindent or equivalently, the probability that a pebble is accreted by the core.  We note that $\epsilon_{PA}=0.1$ is a relatively large value, compared to those derived for low mass discs, which are several orders of magnitude smaller \citep{Liu2018,Ormel2018}.The core's growth is limited by its Type I migration, so the final mass of the planet is given by the mass accretion permitted during the Type I migration timescale.  In other word, by setting the growth timescale equal to the Type I migration timescale, we can determine the final mass of the planet $M_{\rm c}=M_{\rm cross}$.  We follow \citet{Ormel2017} by assuming the Type I migration timescale

\begin{equation}
t_I = \left(\frac{H_g}{r}\right)^2 \frac{M^2_*}{\gamma_I M_{\rm c} \Sigma_g r^2 \Omega}
\end{equation}

\noindent Where $\gamma_I$ is an order of unity constant \citep[cf][]{Kley2012}.  Consequently, a planet that has traversed from $\rsmax$ to $r$ will possess a crossing mass

\begin{equation}
M_{\rm cross}(r) = \sqrt{\frac{3 \pi \alpha |\rsmax-r| \mdotpeb \epsilon_{PA}}{r \mdotgas\gamma_I r }} \left(\frac{H_g}{r}\right)^2 M_*
\end{equation}

\noindent We assume that our embryo ceases growth once it reaches the pebble isolation mass, i.e. where the embryo's Hill radius exceeds the local pebble scale height \citep{Lin1993,Ormel2017}:

\begin{equation}
M_{\rm cross}(r) < M_{\rm iso, peb} = \left(\frac{H_{\rm peb}}{r}\right)^3 M_* \label{eq:pebbleisolationmass}
\end{equation}

We also demand that the total pebble mass in the disc be conserved, i.e. that an embryo cannot accrete more than the available reservoir at its given efficiency.  As the total disc mass $M_{\rm disc}$ is an output of the self-gravitating disc model, we can make the simple constraint

\begin{equation}
M_{\rm cross}(r) < Z_0 f_{\rm peb} M_{\rm disc} \label{eq:mcross_mdisc}
\end{equation}

\section{Results}
\label{sec:results}

\subsection{Pebble Accretion by Disc Fragments}

\begin{figure*}
\begin{center}
\includegraphics[width=0.49\textwidth]{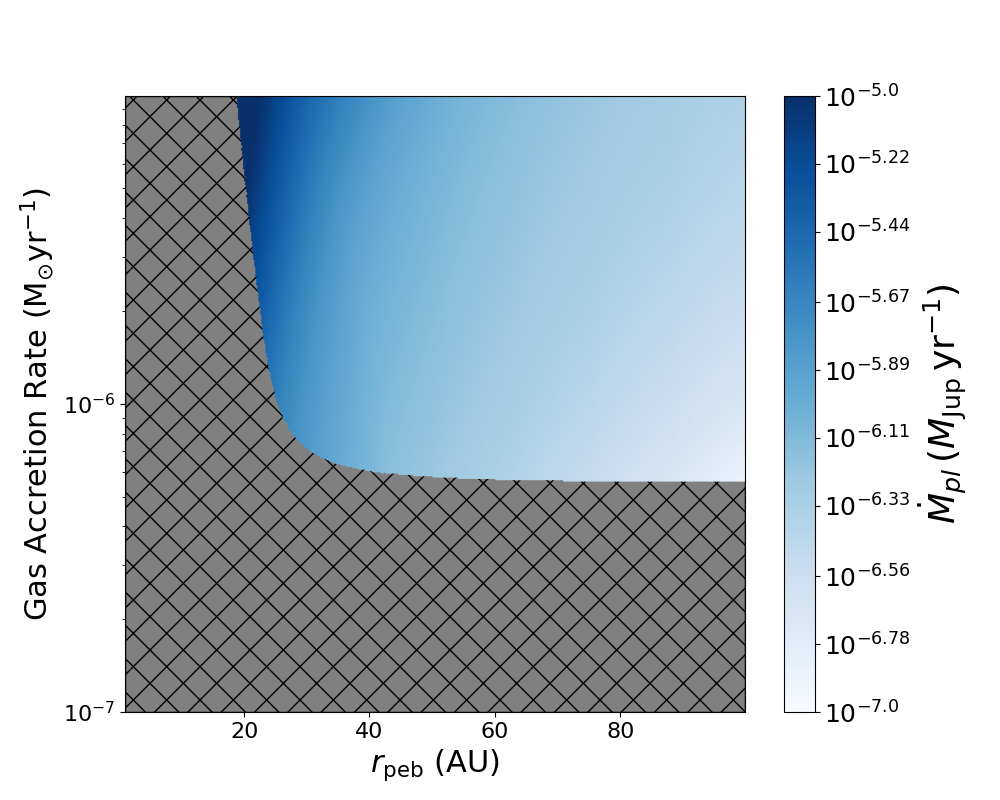} 
\includegraphics[width=0.49\textwidth]{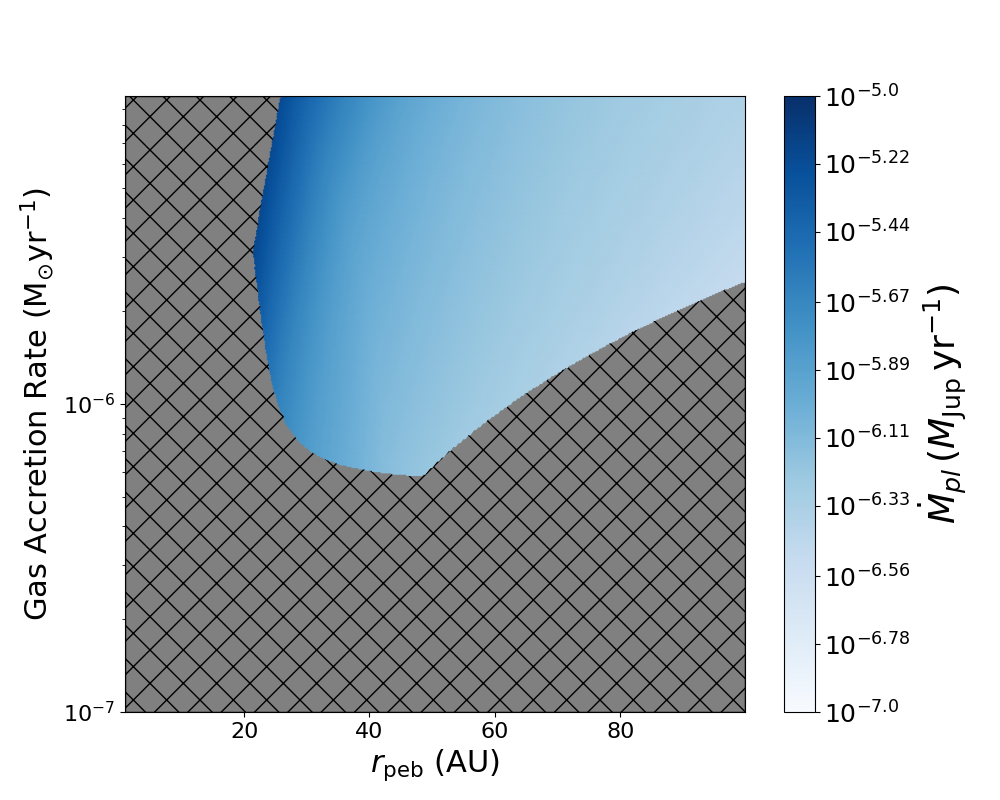}
\end{center}
\caption{Accretion of pebbles by disc fragments.  Left: Pebble accretion assuming strict adherence to the torque balance and gap opening timescale criteria for opening gaps in the gas and pebble discs (equations \ref{eq:gap_torque_balance}, \ref{eq:gap_opening_timescale} and \ref{eq:pebbleisolationmass}).  Right: The same, assuming that shallow gaps can be opened provided the torque balance value $<3$. We find that regardless of input dust population, if no gap is opened the fragments accrete pebbles at the maximal rate (given by the inward pebble accretion flow $\mdotpeb$).  Hashed regions indicate either no fragments present, or completely suppressed pebble accretion due to gaps being opened in both the gas and pebble disc simultaneously. \label{fig:planetmdot}}
\end{figure*}

\noindent We find that regardless of our assumed grain properties (and of pebble disc gap opening, which we find all fragments can easily achieve) all of our fragments can accrete at rates constrained only by the (fixed) pebble accretion efficiency (see Figure \ref{fig:planetmdot}).  While all our fragments exceed the local pebble isolation mass
according to equation \ref{eq:pebbleisolationmass}, and open gaps in the pebble disc, they fail to open gaps in the gas and therefore migrate inwards in the Type I regime.  This ensures that the fragment’s pebble feeding zone continues to be replenished.”  This is in good agreement with numerical simulations of pebble accretion by disc fragments \citep[e.g.][]{Humphries2018}.  The timescale for a fragment to accrete its current mass in pebbles

\begin{equation}
t_{\rm accrete} = \frac{M_{\rm frag}}{\mdotp} \sim 0.1 - 10 \,\rm{Myr}
\end{equation}

\noindent Given that the typical core mass of a disc fragment ranges between 1-10$\mearth$ , this maximal rate would suggest that a core's worth of pebbles could be accreted within $10^2-10^3$ years, which is slightly less than the typical Type I migration timescale.

The fact that any pebble accretion occurs at all is because the fragment cannot open a gap in the gas disc, and therefore migrates inward via Type I.  We find that the torque balance criterion is only weakly satisfied at large disc radii (taking on values $\Gamma_{\rm balance}< 3$), and that $t_{\rm gap} < t_{\rm cross}$ in general, suggesting that fragments are likely to carve shallow gaps.  If we demand $\Gamma_{\rm balance} <1$ for gap opening, then we obtain the results shown in the left panel of Figure \ref{fig:planetmdot}.  If we instead demand $\Gamma_{\rm balance} <3$, we obtain the results showin in the right panel of Figure \ref{fig:planetmdot}.

However, we should take heed of \citet{Malik2015a}'s numerical simulations that suggest gap opening for massive fragments is harder than is apparent from the above set of equations, and that $t_{\rm gap}$ may be increased by a factor as large as 1000.  We find that increasing $t_{\rm gap}$ by a factor of around 3000 completely suppresses gap opening for all disc fragments, and the pebble accretion rate will again match the behaviour seen in the left panel of Figure \ref{fig:planetmdot}.

However, even if gap formation is frustrated by rapid migration, we should also consider the role of turbulence in suppressing local pebble accretion efficiency, both in increasing the relative velocity between pebble and accretor, and in reducing the midplane density of pebbles.  \citet{Liu2018} and \citet{Ormel2018} derive a prescription for $\epsilon_{PA}$for low mass discs with vertically driven turbulence, and find that $\epsilon_{PA}$ can be as low as $10^{-5}$ depending on planet mass, semimajor axis and location of the H$_2$O snowline.  Their findings indicate that such low efficiencies are more typically found at large semimajor axis (i.e 30 au), so this would point towards our initial value of $\epsilon_{PA}=0.1$ being potentially a large over-estimate.  

\citet{Guillot2014} give the eccentricity of a pebble induced by turbulence as

\begin{equation}
e \sim 3 \times 10^{-3} \left(\frac{\alpha}{10^{-3}}\right) \left(\frac{s}{100\,\rm{km}}\right)^{1/3} \left(\frac{r}{1 \rm{au}}\right)^{11/12}
\end{equation}

\noindent suggesting this is only effective for bodies beyond 100 km in size (i.e. planetesimals), in discs where $\alpha > 10^{-3}$.   However, in a self-gravitating disc with $\alpha\sim 0.1$ at around 50 au, metre-sized bodies will achieve eccentricities of order 0.2 \citep[cf][]{Walmswell2013}, although this assumes the body survives beyond the fragmentation limit.  Millimetre grains are expected to have relatively low eccentricities of $e\sim0.02$.  Conversely, $\epsilon_{PA}$ increases with \emph{planet} eccentricity \citep{Liu2018}.  Disc fragment eccentricities are initially rather low at formation: $e<0.1$ \citep{Hall2017}, but can be excited by dynamical interactions \citep{TD_dynamics,Li2016,TD_nbody}. 

\subsection{The Streaming Instability in Self-Gravitating Discs}

\begin{figure*}
\centering
\includegraphics[width=0.49\textwidth]{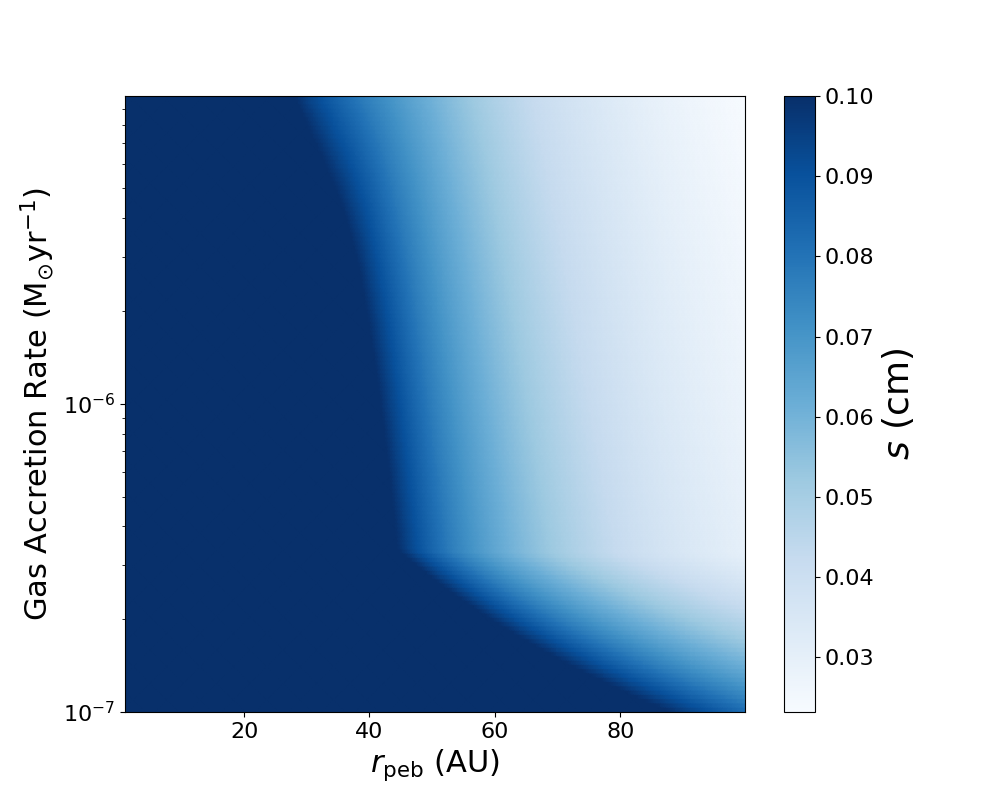}
\includegraphics[width=0.49\textwidth]{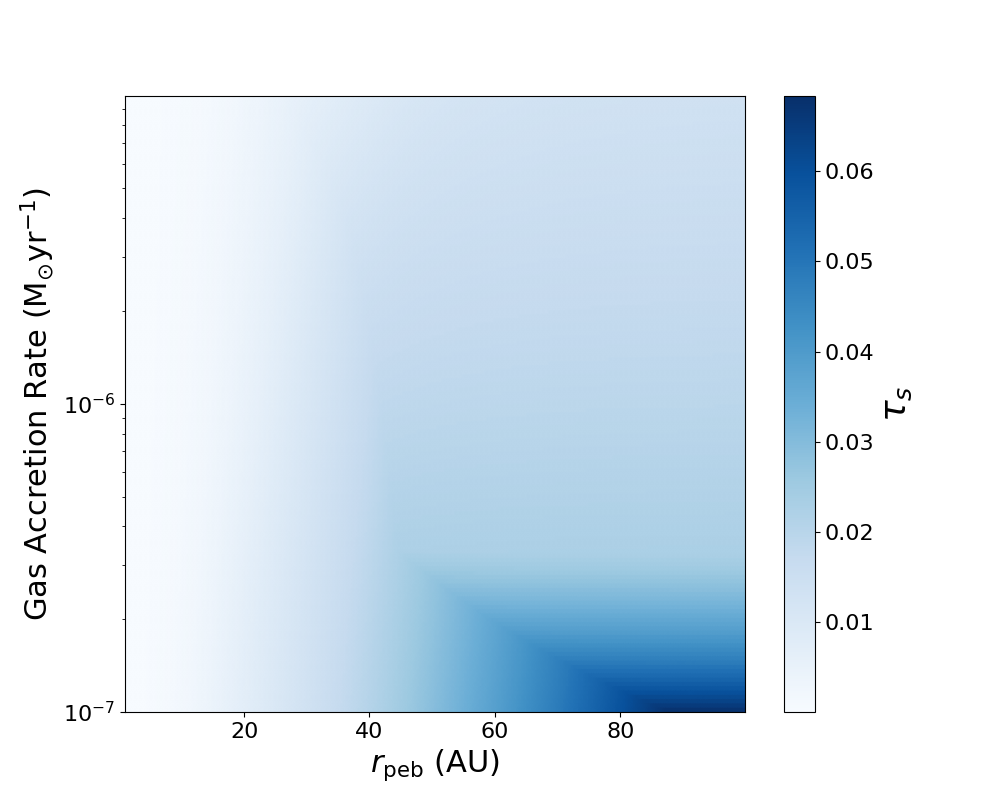}
\includegraphics[width=0.49\textwidth]{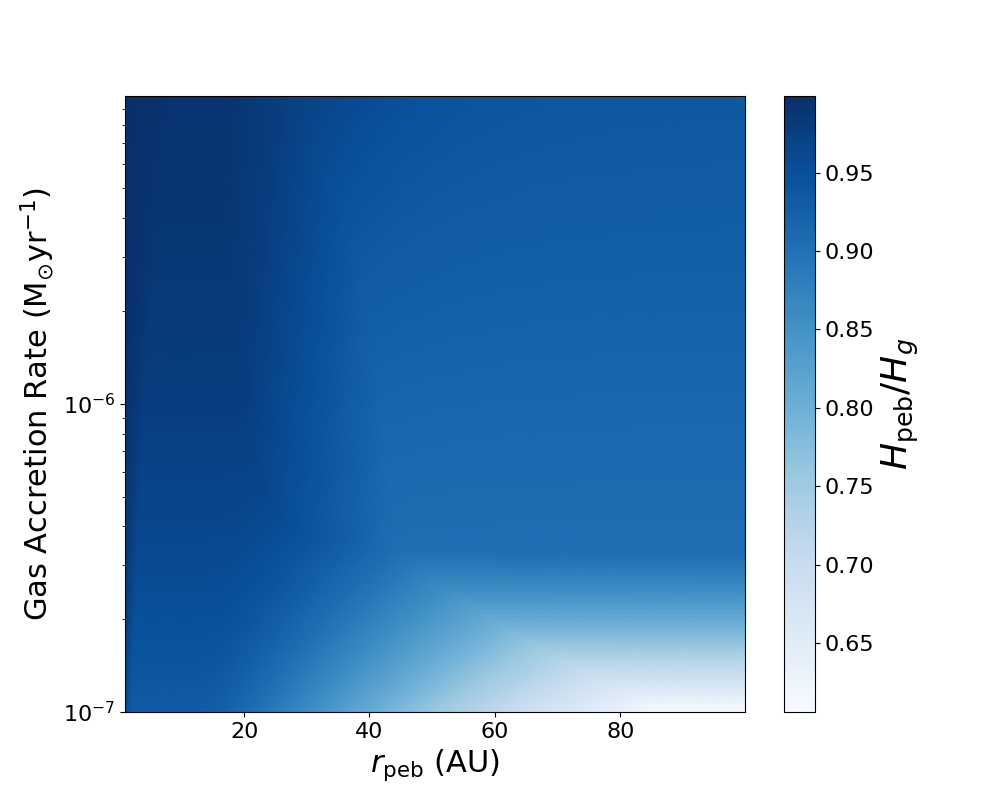}
\includegraphics[width=0.49\textwidth]{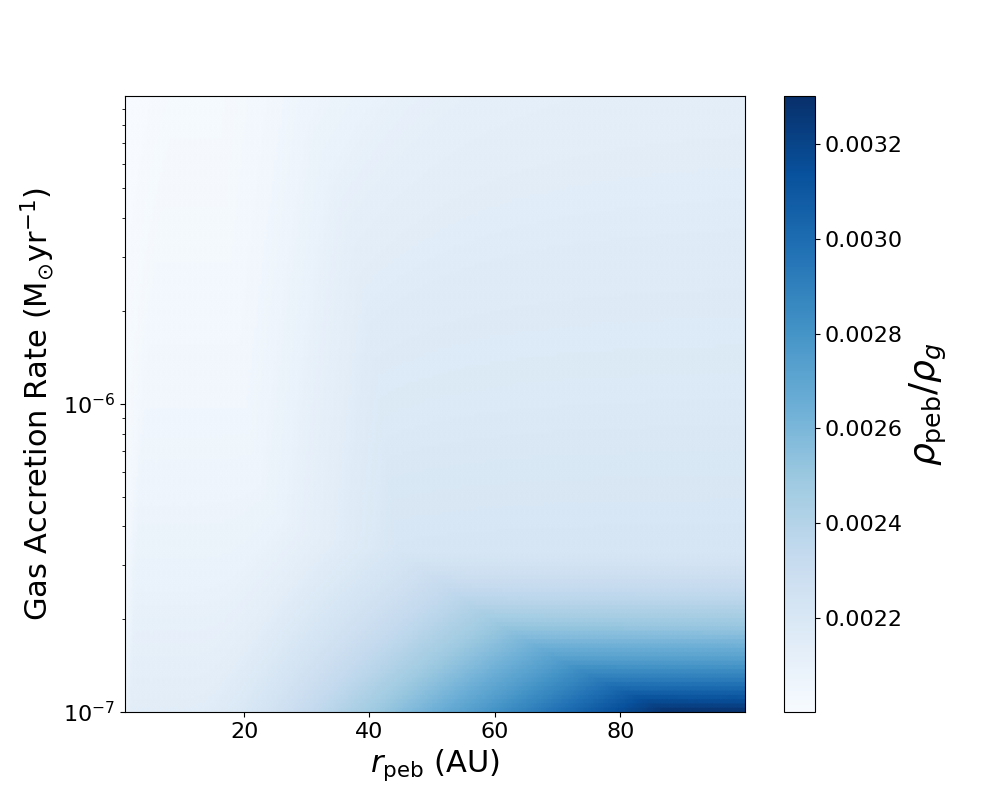}
\caption{Parameters relating to the streaming instability in self-gravitating discs.  Top left: the local grain size $s$ after being restricted by collisional fragmentation (the maximum initial grain size is $s=1$mm).  Top right: the corresponding dimensionless stopping time $\tau_s$.  Bottom left: the ratio of scale heights for the dust and gas discs, $H_{\rm peb}/H_g$.  Bottom right: the ratio of dust volume density to gas volume density, $\rho_p/\rho_g$.}
\label{fig:1mm}
\end{figure*}

In Figure \ref{fig:1mm} we show the results of our calculations of pebble growth (in the absence of disc fragments) with an initial pebble size $s=1$ mm.  The maximum permitted grain size is reduced at larger radius due to collisional fragmentation, which increases the corresponding dimensionless stopping time to maximum values of $\tstop\approx 0.06$.  Note that this effectively rules out enhanced grain growth due to concentration in spiral arms, as this process appears to require $\tstop > 0.1$, and hence any concentration here will be modest at best \citep{Booth2016}.

As the pebble disc possesses a relatively large scale height $H_{\rm peb}$ (at minimum around 60\% of the gas scale height $H_g$, bottom left panel of Figure \ref{fig:1mm}), and the available pebble supply is restricted by the local dust mass reservoir, we find that $\rho_p/\rho_g \approx 10^{-3}$ for our parameter space (bottom right panel of Figure \ref{fig:1mm}).   As a result, the streaming instability fails to generate planetesimals.  This is consistent with more advanced models of disc formation and planetesimal growth, such as those recently presented by \citet{Drazkowska2018}. 

We attempted to activate the streaming instability by modifying the input parameters.  We found that by increasing the grain size to $s=10$ cm, enhancing the individual grain density to $\rho_s = 5 \,\rm{g\,cm^{-3}}$, setting $f_{\rm peb}=1$ and $Z=0.04$, the inner disc (1-2 au) becomes streaming unstable, and forms planetesimals with a crossing mass of $M_{\rm cross}\sim 10^{-2}-10^{-3} \mearth$, i.e. of order a lunar mass.  Of course, these parameters are unphysical, and gravitational instability is likely to be a sub-dominant process in the inner disc, as we expect $Q>2$ for realistic discs.

That being said, if cm-sized grains or larger are present in the disc, then spiral arm concentration can play a more important role, especially in setting the grain radial velocity.  Also, simulations have indicated that disc regions with low turbulence can activate the streaming instability even for $\rho/\rho_g \sim 10^{-2}$ \citep{Bai2010, Drazkowska2014,Carrera2015}.  As the gravitational stress decreases with proximity to the star, it may well be the case that the inner disc is more suitable for planetesimal formation than our work indicates, especially if spiral structure can assist.

If even a small population of lunar-mass planetesimals can be formed at early times, this skewing of the grain size distribution has important consequences for future planet formation by core accretion.  We therefore conclude that while streaming instability is not likely in young self-gravitating protostellar discs, the possibility of weak activation of the instability within or around spiral structure demands further investigation.

\section{Discussion}
\label{sec:discussion}

\subsection{Implications for objects formed by disc instability}

\noindent These models indicate that provided mm-sized grains are present at early times, disc fragments are likely to accrete them at a substantial rate, with some exceptions in the outer regions where shallow gaps may be opened.  The accretion luminosity resulting from this is likely to provide substantial cooling to the fragment, which will assist in its collapse to form a bound object \citep{Nayakshin2014b}.   

Accreting the majority of the core mass beyond 30 AU at early times may assist in the formation of super-Jupiters at wide separations, such as the HR 8799 system.  It is commonly assumed HR 8799's formation is best explained purely by GI, but building dynamically stable multi-planet systems of this type is not a typical outcome of GI \citep{TD_nbody}, meaning that this explanation is at some level incomplete \citep{Nero2009,Helled2010,Marois2010}.  Indeed, it may be further challenged by indirect evidence for an additional fifth planet \citep{Read2018}. We speculate that pebble accretion may in fact play a crucial role in growing disc fragments and driving gaps, and therefore assembling resonant, wide-orbit multi-planet systems such as HR 8799 via GI.

If the fragment fails to assume a fully bound state before being tidally disrupted, these grains will be returned to the disc, with the distribution of grain size heavily affected by when the grains are accreted, and the sedimentation of said grains towards the centre of the fragment.

Even if pebble accretion is not particularly efficient, these fragments may begin to collect a substantial population of ``near-miss'' pebbles.  Given that most fragments contain significant spin angular momentum \citep{Galvagni2012,Hall2017}, we could expect that these fragments may form circumfragmentary discs that contain an enhanced dust-to-gas ratio relative to the parent protostellar disc.  The formation of satellites around massive bodies could be substantially enhanced by this effect.  A possible example of this process in action is the exomoon candidate Kepler-1625b-i, currently believed to be a satellite several Earth radii in size, orbiting a 10 $\mjup$ planetary mass object \citep[see][]{Teachey2017,Heller2017c,Teachey2018}.

Fragments can also undergo impacts and mergers with neighbour fragments \citep{Hall2017}.  If one pebble-rich fragment is absorbed by a less pebble-rich fragment, then we can expect the resulting change in opacity to have significant effects on the resulting fragment's evolution.  Partial mergers or captures can result in exchange of solids, modifying the opacity of both objects, potentially resulting in the formation of binary planets, with evolutionary trajectories akin to common envelope binary stars.

This all points to the evolution of pebble-rich disc fragments under tidal downsizing to be a highly non-trivial process, especially if we consider dynamical interactions with neighbouring fragments \citep{TD_dynamics,Li2016a,TD_nbody}.


\subsection{Implications for planets formed via core accretion}

\noindent The dynamics of pebble accretion during the disc instability phase is likely to have significant effects on later core accretion.  If the disc fragments, then these fragments will accrete large quantities of the pebble reservoir, either sequestering them in the core of a giant planet, releasing them into the disc during a tidal disruption event, or funneling them onto the central star.  This may result in significant pebble depletion (especially in the outer regions of the disc).  Subsequent attempts to form planets via core accretion will be limited by this depletion.  

If we attempt to model the core accretion of a system where disc fragments have been present, it is likely we will have to invoke a low pebble fraction $f_{\rm peb}$ to explain the observed planets.  If GI is a relatively infrequent process, as some have concluded \citep{Vigan2017}, then one may be able to identify at a statistical level a sub-population of planetary systems that appear to have formed in a pebble-poor environment.  Barring other explanations for a low pebble production rate, this sub-population may potentially indicate the past presence of disc fragments, even if the fragments themselves do not survive.

It could be argued that a similar effect may be produced by a giant planet formed via core accretion, but it is worth noting that a disc fragment will traverse the disc much earlier in its evolution, when most of the solid population remains in grains/pebbles.  In contrast, a giant planet formed via core accretion will migrate inwards through a much more evolved population, potentially hosting protoplanet cores.  Essentially, a disc fragment will accrete/rearrange dust, while a core accretion planet will scatter protoplanets.  These should arguably leave distinct observational signatures in the final planetary system architecture.

If the disc does not fragment, then this pebble depletion will not occur, and there may be some \emph{rare} circumstances in which spiral structure seeds the formation of lunar-mass planetesimals in a still self-gravitating disc.  The fate of these bodies is unclear - they may continue to accrete dust and pebbles to form larger cores, eventually being accreted onto the star thanks to Type I migration. 

Of course, planetesimal impacts can also cause shattering without subsequent re-accretion. This could inject new grain populations into regions of the disc favourable to core accretion.  Any and all of the above effects can rewrite the grain distribution, both in size and spatial distribution.  The range of permitted initial conditions for core accretion may be much larger than previously thought, with some quite non-trivial distribution functions for grain parameters.  

This undoubtedly propagates into the structural and orbital parameters of planets formed via core accretion.  Careful modelling of this propagation may yield important clues to the early history of the system, in particular precisely how the processes of disc instability and core accretion interact, which remains an important (and poorly understood) constraint for any final theory of exoplanet/brown dwarf formation.

\section{Conclusions}
\label{sec:conclusions}

\noindent In this paper, we have applied pebble accretion theory to steady-state models of gravitationally unstable protostellar discs.  We find that if these discs are prone to fragmentation, disc fragments can accrete mm-sized bodies vigorously, at a rate limited principally by the local pebble accretion efficiency (and the disc's supply of pebbles).  This may assist the survival of these objects through effective cooling, and/or result in rapid injection of planetesimals back into the disc when fragments undergo tidal disruption.

In general, we find the streaming instability fails to operate amongst mm grains.  If larger grains exist, concentration of the grains in spiral structure may activate the instability, but this remains speculation at best.  If this occurs, lunar-mass planetesimals can potentially form in relatively young self-gravitating discs.  This will clearly have effects on the resulting grain and planetesimal populations, which at this moment are difficult to quantify, but are clearly very different from those commonly used in core accretion models of planet formation.  

In any case, if disc fragments form, their inward migration and accretion of pebbles will reconfigure the spatial and size distributions of solid bodies that can then participate in core accretion.  In other words, pebble accretion creates small but important linkages between gravitational instability and core accretion models of planet formation, and these linkages must be understood.

We admit that our conclusions make assumptions that are not yet borne out by appropriate simulations of mm-sized and cm-sized grains growing in gravito-turbulence.  In particular, it is unclear how the streaming instability or something equivalent operates under relatively strong turbulent stresses in self-gravitating discs, or what the pebble accretion efficiency is likely to be in this gravito-turbulent state.  We advocate for further research in these areas, which will clarify how the self-gravitating phase of protostellar discs rewrites the initial conditions for planet formation.



\section*{Acknowledgements}

The author gratefully acknowledges support from the ECOGAL project, grant agreement 291227, funded by the European Research Council (ERC) under ERC-2011-ADG.   This manuscript benefitted greatly from conversations with Richard Booth and Farzana Meru, and a very attentive reading by the anonymous reviewer.  This research has made use of NASA's Astrophysics  Data System Bibliographic Services.  The code used in this paper is available at \url{https://github.com/dh4gan/sgd-grid}




\bibliographystyle{mnras} 
\bibliography{sgd_pebbles}


\bsp	
\label{lastpage}
\end{document}